\documentclass[journal]{IEEEtran}
\usepackage{cite}
\usepackage[pdftex]{graphicx}
\DeclareGraphicsExtensions{.pdf,.jpeg,.png,.jpg}
\usepackage[cmex10]{amsmath}
\usepackage{bigstrut}
\usepackage{multirow}

\usepackage{times}
\usepackage{graphicx}
\usepackage{amsmath}
\usepackage{lettrine}
\usepackage{setspace}
\usepackage{epstopdf}

\usepackage{url}

\begin{document}

\title{Design Exploration of Hybrid CMOS-OxRAM Deep Generative Architectures}

\author{\IEEEauthorblockN{Vivek Parmar,~\IEEEmembership{Member,~IEEE,} and Manan Suri,~\IEEEmembership{Member,~IEEE}}
\\\IEEEauthorblockA{Indian Institute of Technology-Delhi, Hauz Khas, New Delhi-110016,
email: manansuri@ee.iitd.ac.in}}

\maketitle

\begin{abstract}
Deep Learning and its applications have gained tremendous interest recently in both academia and industry. Restricted Boltzmann Machines (RBMs) offer a key methodology to implement deep learning paradigms. This paper presents a novel approach for realizing hybrid CMOS-OxRAM based deep generative models (DGM). In our proposed hybrid DGM architectures, HfOx based (filamentary-type switching) OxRAM devices are extensively used for realizing multiple computational and non-computational functions such as: (i) Synapses (weights), (ii) internal neuron-state storage, (iii) stochastic neuron activation and (iv) programmable signal normalization. To validate the proposed scheme we have simulated two different architectures: (i) Deep Belief Network (DBN) and (ii) Stacked Denoising Autoencoder for classification and reconstruction of hand-written digits from a reduced MNIST dataset of 6000 images. Contrastive-divergence (CD) specially optimized for OxRAM devices was used to drive the synaptic weight update mechanism of each layer in the network. Overall learning rule was based on greedy-layer wise learning with no back propagation which allows the network to be trained to a good pre-training stage. Performance of the simulated hybrid CMOS-RRAM DGM model matches closely with software based model for a 2-layers deep network. Top-3 test accuracy achieved by the DBN was 95.5\%. MSE of the SDA network was 0.003, lower than software based approach. Endurance analysis of the simulated architectures show that for 200 epochs of training (single RBM layer), maximum switching events/per OxRAM device was $\sim$ 7000 cycles. 

\end{abstract}




%
\IEEEpeerreviewmaketitle

\section{Introduction}
\IEEEPARstart{T}{he} neurons in our brain have a capacity to process a large amount of high dimensional data from various sensory inputs while still focusing on the most relevant components for decision making \cite{pillow2006dimensionality} \cite{cunningham2014dimensionality}. This implies that the biological neural networks have a capacity to perform dimensionality reduction to facilitate decision making. In the field of machine learning, artificial neural networks also require a similar capability because of the availability of massive amounts of high dimensional data being generated everyday through various sources for digital information. Thus it becomes imperative to derive an efficient method for dimensionality reduction to facilitate tasks like classification, feature learning, storage, etc. Deep generative networks such as Autoencoders have been shown to perform better than many commonly used statistical techniques such as PCA (principal component analysis), ICA (Independent Component Analysis) for encoding and decoding of high dimensional data \cite{hinton1994autoencoders}. These networks are traditionally trained using gradient descent based on back-propagation. However it is observed that for deep networks, gradient descent doesn't converge and gets stuck in a local minima in case of purely randomized initialization \cite{bengio2007greedy}. A solution to this problem is, weight initialization by utilizing a generative layer-by-layer training procedure based on Contrastive Divergence (CD) algorithm \cite{hot06}.

To maximize the performance of this algorithm, a dedicated hardware implementation is required to accelerate computation speed. Traditionally CMOS based designs have been used for this by utilizing commonly available accelerator like GPUs \cite{raina2009large}, FPGAs \cite{kim2009highly}, ASICs \cite{maaimm11}\cite{stromatias2015robustness}, etc. Recently with the introduction of the emerging non-volatile memory devices such as PCM, CBRAM, OxRAM, MRAM, etc, there is further optimization possible in design of a dedicated hardware accelerators given the fact that they allow replacement of certain large CMOS blocks while simultaneously emulating storage and compute functionalities \cite{sqcpsvgd11}\cite{alibart2013pattern}\cite{yang2013memristive}\cite{de2013silicon}\cite{jackson2013nanoscale}\cite{vlzrbgkgq14},\cite{wong2015memory}\cite{burr2015experimental}\cite{milo2016demonstration}.  

Recent works that present designs of Contrastive Divergence based learning using resistive memory devices are \cite{srpj15}, \cite{stanford_rbm_prob}. In \cite{srpj15} the authors propose the use of a two-memristor model as a synapse to store one synaptic weight. In \cite{stanford_rbm_prob} the authors have experimentally demonstrated a 45-synapse RBM realized with 90 resistive phase change memory (PCM) elements trained with a bio-inspired variant of the contrastive divergence algorithm, implementing Hebbian and anti-Hebbian weight update. Both these designs justify the use of RRAM devices as dense non-volatile synaptic arrays. Also both make use of a spike based programming mechanism for gradually tuning the weights. Negative weights have been implemented by using two devices in place of a single device per synapse. It is apparent that in order to implement more complex learning rules with larger and deeper networks the hardware complexity and area footprint increases considerably while using this simplistic design strategy. As a result, there is a need to increase further increase the functionality of the RRAM devices in the design beyond simple synaptic weight storage. In \cite{tnanoelm} we have described a design exploiting the intrinsic device-to-device variability as a substitute for the randomly distributed hidden layer weights in order to gain both area and power savings. In \cite{rramrbm}, we have made use of another property of the RRAM devices by exploiting the cycle-to-cycle variability in device switching to create a stochastic neuron as a basic building block for a hybrid CMOS-OxRAM based Restricted Boltzmann Machines (RBM) circuit.   




In this paper we build upon our previous work on hybrid CMOS-OxRAM RBM with the following novel contributions: 
\begin{itemize}
\item
Design of deep generative models (DGM) that utilize the hybrid CMOS-RRAM RBM as a building block.
\item
Design of programmable output normalization block for stacking multiple hybrid RBMs.
\item
Simulation and performance analysis of two types of DGM architectures at 8-bit synaptic weight resolution: (i) Deep Belief Networks (DBN) and (ii) Stacked Denoising Autoencoders (SDA) \item
Analysis of learning performance (accuracy, MSE) while using only greedy layer-wise training (without backprop).
\item
Analysis of learning impact on RRAM device endurance. 
\end{itemize}
In our hybrid CMOS-OxRAM DGM implementation the OxRAM devices have been exploited for four different storage and compute functions: (i) Synaptic weight matrix, (ii) neuron internal state storage, (iii) stochastic neuron firing and (iv) programmable gain control block. 
Section \ref{s2} discusses the basics of OxRAM and deep generative networks. Section \ref{s4} describes the implementation details of our proposed hybrid CMOS-OxRAM DGM architectures. Section \ref{s5} discusses simulation results and Section \ref{sc} gives the conclusions.

\section{Basics of OxRAM and DGM Architectures}
\label{s2}
\subsection{OxRAM Working}
\begin{figure}[htbp]
\centering
\includegraphics[scale=.42]{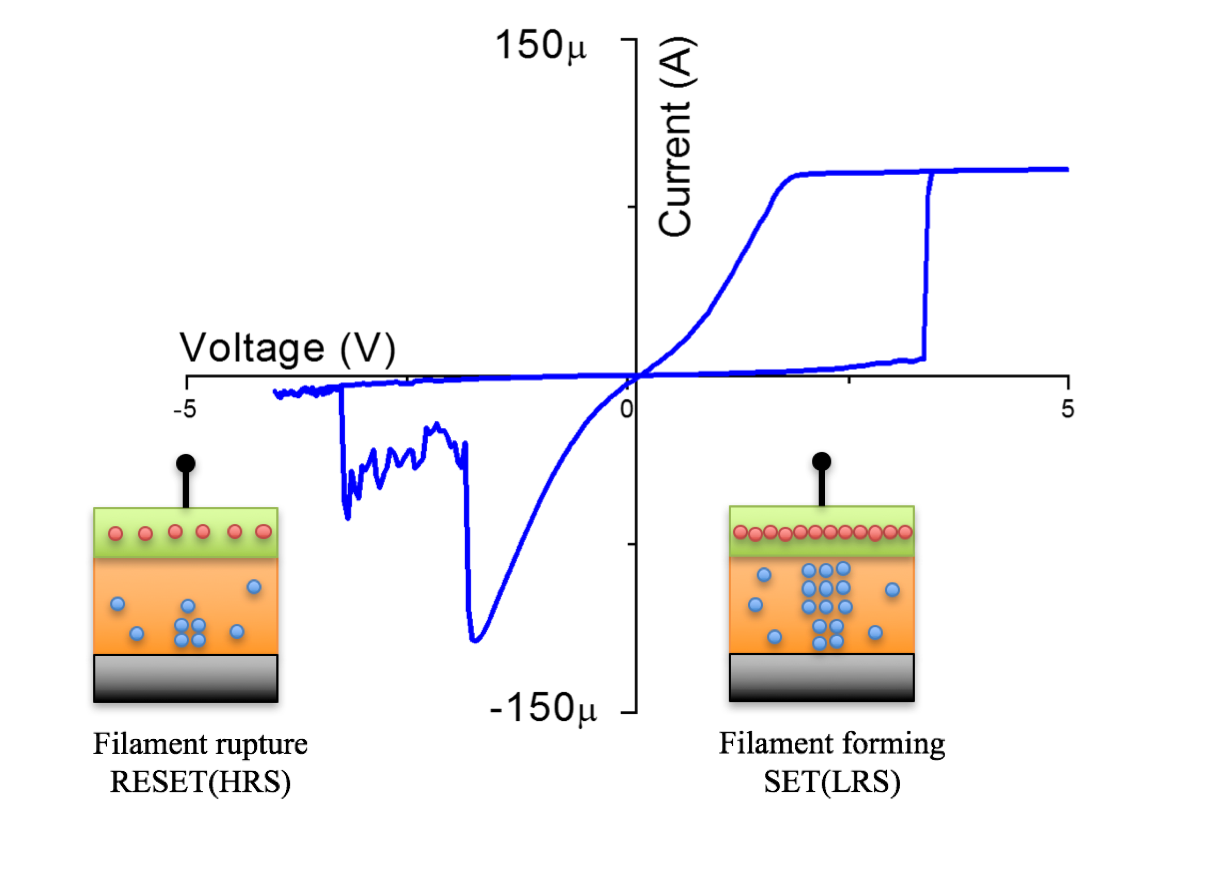}
\caption{Basic IV characteristics for HfOx OxRAM device with switching principle indicated. Experimental data corresponding to device presented in \cite{rramrbm}.}
\label{fig:1a}
\end{figure}
\begin{figure}[htbp]
\centering
\includegraphics[scale=.3]{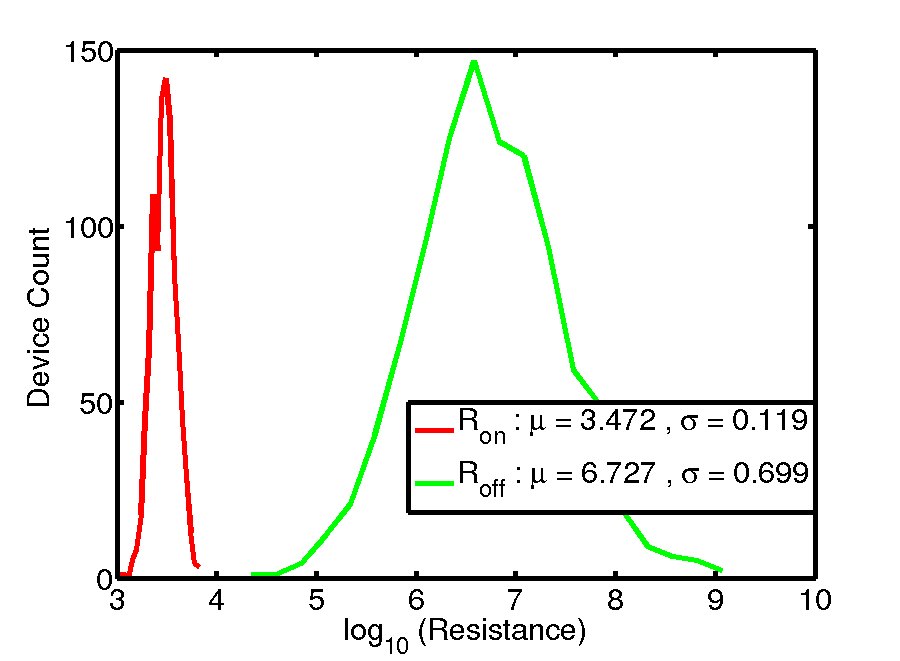}
\caption{Cycle-to-Cycle ON/OFF-state resistance distribution for HfOx device presented in \cite{rramrbm}.}
\label{fig:1b}
\end{figure}
OxRAM devices are two-terminal MIM-type structures (metal-insulator-metal) sandwiching an active metal-oxide based insulator layer, between metallic electrodes (see Fig. \ref{fig:1a}). The active layer exhibits reversible non-volatile switching behavior on application of appropriate programming current/voltage across the device terminals. In the case of filamentary- OxRAM devices, formation of a conductive filament in the active layer, leads the device to a low-resistance (LRS/On) SET-state, while dissolution of the filament puts the device in a high-resistance (HRS/Off) RESET-state. The conductive filament is composed of oxygen vacancies and defects \cite{wlycwclct12}. SET-state resistance (LRS) level can be defined by controlling the dimensions of the conductive filament \cite{sqbpvvgd13}\cite{wlycwclct12}, which depends on the amount of current flowing through the active layer. Current flowing through the active layer is controlled either by externally imposed current compliance or by using an optional selector device (i.e. 1R-1T/1D configuration). 
OxRAM devices are known to demonstrate cycle-to-cycle (C2C) (shown in Fig. \ref{fig:1b}), and device-to-device (D2D) variability \cite{baeumer2017subfilamentary}\cite{li2015variation},\cite{ielmini2016resistive}. In our proposed architecture, we exploit OxRAM (a) C2C switching variability for realization of stochastic neuron circuit, (b) binary resistive switching for realization of synaptic weight arrays/neuron internal state storage and (c) SET-state resistance modulation for normalization block.

\label{s3}
\subsection{Restricted Boltzmann Machines (RBM)}
Unsupervised learning based on generative models has gained importance with use of deep neural networks. Besides being useful for pre-training a supervised predictor, unsupervised learning in deep architectures can be of interest to learn a distribution and generate samples from it \cite{bengio2009learning}. 
RBMs in particular, are widely used as building blocks for deep generative models such as DBN and SDA. Both these models are made by stacking RBM blocks on top of each other. Training of such models using traditional back-propagation based approaches is a computationally intensive problem. Hinton et.al.\cite{hot06} showed that such models can be trained very fast through greedy layer-wise training making the task of training deep networks based on stacking of RBMs more feasible.
\begin{figure}[b]
\centering
  \includegraphics[scale=0.5]{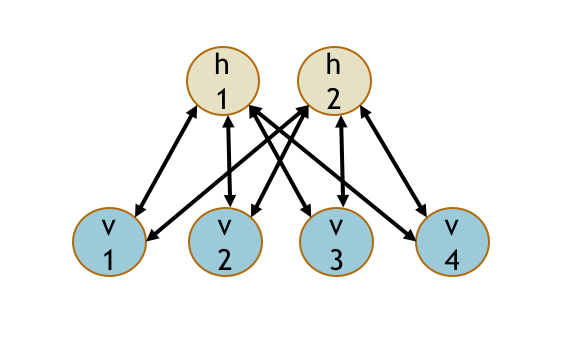}
  \caption{Graphical representation of RBM hidden/visible nodes.}
  \label{fig1a}
\end{figure}
Each RBM block consists of two layers of fully connected stochastic sigmoid neurons as shown in Fig. \ref{fig1a}. The input or the first layer of the RBM is called the visible layer and the second (feature detector) layer is called the hidden layer. Each RBM is trained using CD algorithm as described in \cite{h10}. The output layer of the bottom RBM acts as the visible layer for the next RBM.

\subsection{Stacked Denoising Autoencoder (SDA)}
An autoencoder network is a deep learning framework mostly used for denoising corrupted data \cite{vincent}, dimensionality reduction \cite{hinton1994autoencoders} and weight initialization applications. In recent years random weight initialization techniques have been preferred over use of generative training networks \cite{xavier}, however DGMs continue to be the ideal candidate for dimensionality reduction and denoising applications. 
Autoencoder network is basically realized using two networks:
\begin{enumerate}
  \item An 'encoder' network which has layers of RBMs stacked on the top of one another.
  \item A mirrored 'decoder' network with same weights as that of the encoder layer for data reconstruction.
\end{enumerate}

\begin{figure}[b]
\centering
  \includegraphics[width=\linewidth]{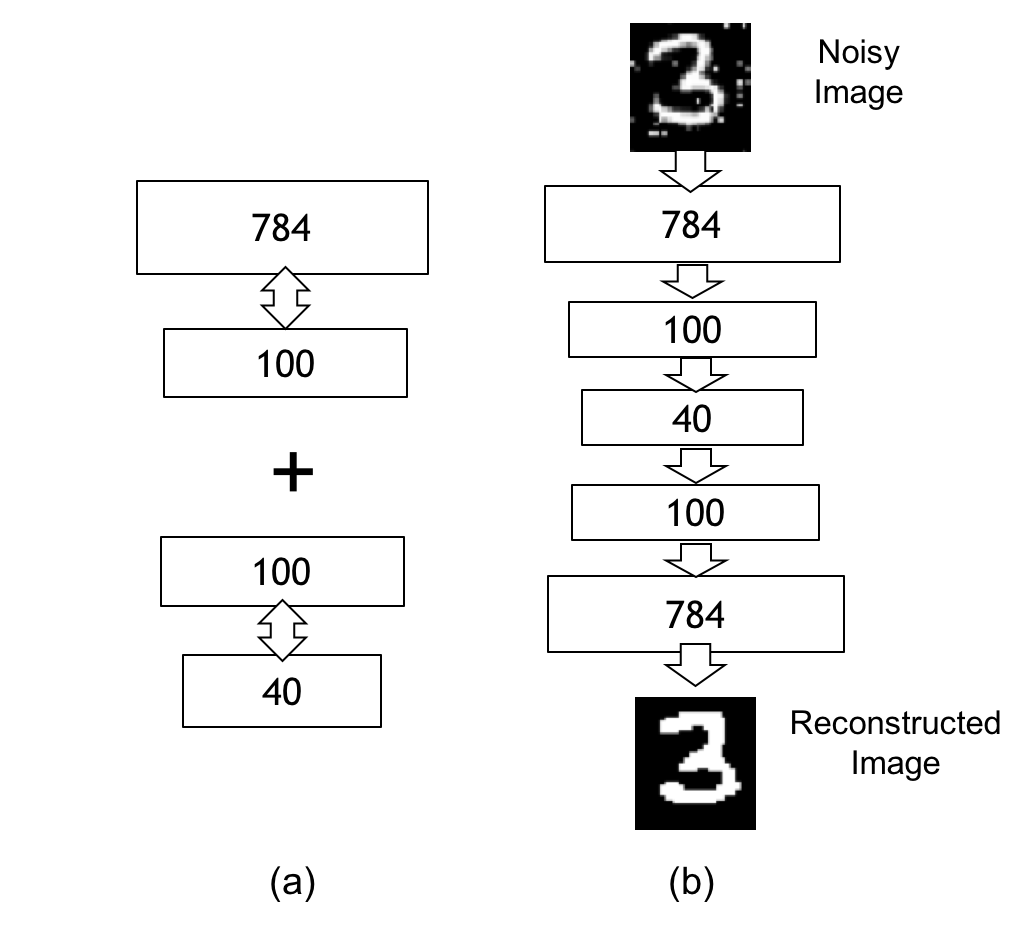}
  \caption{(a) Basic RBM blocks stacked to form a deep autoencoder. (b) Denoising noisy image using autoencoder.}
  \label{fig1}
\end{figure}
The stack of RBMs in autoencoder are trained layer-wise one after the other. An 'unrolled' autoencoder network with the encoder and decoder is shown in Fig. \ref{fig1}. 

\subsection{Deep Belief Network (DBN)}
\begin{figure}[b]
\centering
  \includegraphics[width=0.7\linewidth]{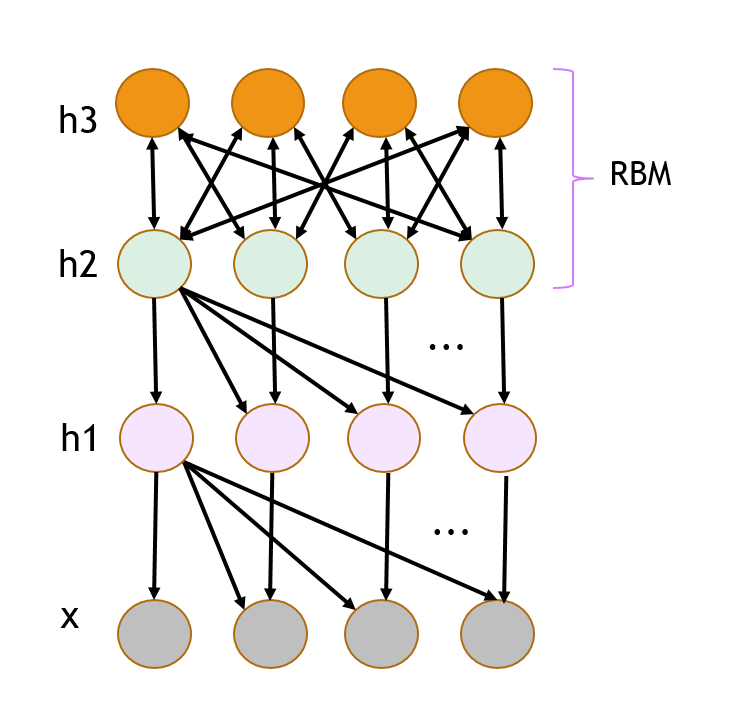}
  \caption{DBN architecture comprising of stacked RBMs}
  \label{fig1b}
\end{figure}

DBNs are probabilistic generative models that are composed of multiple layers of stochastic, latent variables \cite{hinton2009deep}. The latent variables typically have binary values and are often called hidden units or feature detectors. The top two layers have undirected, symmetric connections between them and form an associative memory. The lower layers receive top-down, directed connections from the layer above. The states of the units in the lowest layer represent a data vector. A typical DBN is shown in Fig. \ref{fig1b} which uses a single RBM as the first two layers followed by a sigmoid belief network (logistic regression layer) for the final classification output.

The two most significant DBN properties are:
\begin{enumerate}
\item There is an efficient, layer-by-layer procedure for learning the top-down, generative weights that determine how the variables in one layer depend on the variables in the layer above.
\item After learning, the values of the latent variables in every layer can be inferred by a single, bottom-up pass that starts with an observed data vector in the bottom layer and uses the generative weights in the reverse direction.
\end{enumerate}
DBNs have been used for generating and recognizing images \cite{hot06}, \cite{huang2007unsupervised},\cite{bengio2007greedy}, video sequences \cite{sutskever2007learning}, and motion-capture data \cite{taylor2007modeling}. With low number of units in the highest layer, DBNs perform non-linear dimensionality reduction and can learn short binary codes, allowing very fast retrieval of documents or images \cite{hinton2006reducing}.

\section{Implementation of Proposed Architectures}
\label{s4} 
\begin{figure*}[htbp]
  \includegraphics[width=\linewidth]{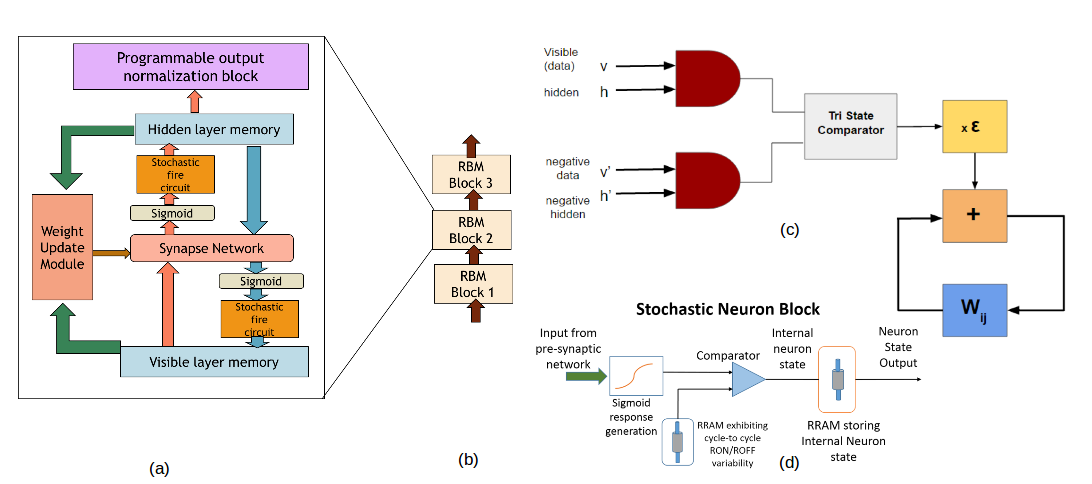}
  \caption{(a) Individual RBM training layer architecture. RBM training block symbols, 'H', 'V' and 'S' represent hidden layer memory, visible layer memory, and synaptic network respectively. (b) Cascaded RBM blocks for realizing the proposed deep autoencoder with shared weight update module (c) Fully digital CD based weight update module. (d) Block level design of single stochastic sigmoid neuron}
  \label{fig2}
\end{figure*}
Basic building block of both SDA and DBN is the RBM. In our simulated architectures, within a single RBM block OxRAM devices are used for multiple functionalities. The basic RBM block (shown in Fig. \ref{fig2}(a)) is replicated, with the hidden layer memory states of the first RBM acting as visible layer memory for the next RBM block and so on (Fig. \ref{fig2}(b)). All RBM blocks have a common weight update module described in Section \ref{CDup}. Post training, the learned synaptic weights along with the sigmoid block can be used for reconstructing the test data. Architecture sub-blocks consist of:

\subsection{Synaptic Network}
Synaptic network of each RBM block was simulated using a 1T-1R HfOx OxRAM matrix. Each synaptic weight is digitally encoded in a  group of binary switching OxRAM devices, where the number of devices used per synapse depends on the required weight resolution. For all architectures simulated in this work we have used 8-bit resolution (i.e. 8 OxRAM devices/per synapse).

\subsection{Stochastic Neuron Block}

Fig \ref{fig2}(d), shows the stochastic sigmoid neuron block. Each neuron (hidden or visible) has a sigmoid response, which was implemented using a low-power 6-T sigmoid circuit (\cite{suri5}). Gain of the sigmoid circuit can be tuned by optimizing the scaling of the six transistors. Voltage output of the sigmoid circuit is compared with the voltage drop across the OxRAM device, with the help of a comparator. The HfOx based device is repeatedly cycled ON/OFF. C2C intrinsic $R_{ON}$ and $R_{OFF}$ variability of the OxRAM device leads to a variable reference voltage for the comparator. This helps to translate the deterministic sigmoid output to a neuron output, which is effectively stochastic in nature. At any given moment, a specific neuron's output determines it's internal state, which needs to be stored for RBM driven learning. Neuron internal state is stored using individual OxRAM devices placed after the comparator. Single OxRAM/per neuron is sufficient for state storage, since RBM requires each neuron to only have a binary activation state.

\subsection{CD Weight Update Block}
\label{CDup}
The weight update module is a purely digital circuit that reads the synaptic weights and internal neuron states. It updates the synaptic weights during learning based on the CD RBM algorithm (\cite{h10}). The block consists of an array of weight update circuits, one of which is shown in Fig \ref{fig2}(c). Synaptic weight is updated by ${\bigtriangleup}$Wij \ref{eqcd}, based on the previous (v, h) and current (v’, h’) internal neuron states of the mutually connected neurons in the hidden and visible layers. CD is is realized using two AND gates and a comparator (having outputs -1, 0, +1). Input to the first AND gate is previous internal neuron states, while the input to second AND gate is the current internal neuron states. Based on the comparator output, ${\epsilon}$ (learning rate) will either be added, subtracted, or not applied to the current synaptic weight (Wij). 

\begin{equation}
\label{eqcd}
\bigtriangleup W_{ij}= \epsilon (vh^T-v^{'}h^{'T})
\end{equation}
\subsection{Output Normalization block}
\begin{figure*}[htbp]
\centering
  \includegraphics[width=\textwidth]{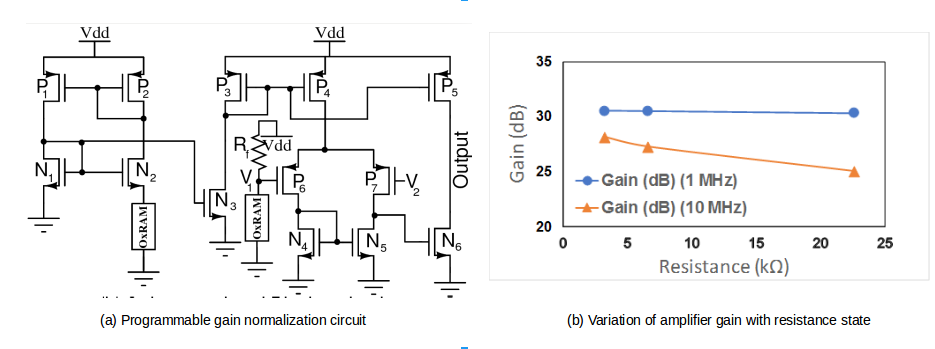}
  \caption{Programmable normalization circuit: (a) Circuit Schematic, (b) Gain variation w.r.to variation in OxRAM resistance state}
  \label{fig_norm}
\end{figure*}

In order to chain the mixed-signal design of RBM we need to ensure the signal output at each layer is having an enhancement in the dynamic range so that the signal doesn't deteriorate as the network depth increases. For this purpose we proposed a hybrid CMOS-OxRAM programmable normalization circuit (see Fig. \ref{fig_norm}) whose gain and bias can be tuned based on OxRAM resistance programming.

The circuit schematic of the programmable normalization block is shown in Fig. \ref{fig_norm}(a)). In order to check variation in gain, we have considered programming the OxRAM in three different SET states ($\sim$ 3.2 k$\Omega$, 6.6 k$\Omega$, and 22.6 k$\Omega$).

The differential amplifier consisting of a DC gain control circuit and a biasing circuit is used to implement the normalization function. A two stage amplifier consisting of transistors N3, N4, N5, N6, P3, P4, P5, P6, and P7 is used. DC gain of the circuit is controlled using a constant $g_m$ circuit whose output is fed into N3. The constant $g_m$ circuit consists of transistors N1, N2, P1, P2 and one OxRAM. Based on the OxRAM resistance, $g_m$ of the circuit can be changed thereby changing the output potential. This affects $V_{gs}$ of N3 thereby controlling the gain of the circuit. 

To validate the design, we performed simulation of the circuit using an OxRAM device compact model (\cite{li2015variation}) and 90 nm CMOS design kit. The simulated variation in the gain of the circuit based on the resistance state of the OxRAM is shown in Fig. \ref{fig_norm}(b). Gain control through OxRAM programming was found to be more prominent at higher operating frequencies. 

Bias control is implemented by a potential divider circuit ($R_f$ and the OxRAM). The potential divider circuit determines the potential across $V_g$ of P6. Input $V_2$ is swept from 0 V to 1 V. If the potential across P6 increases for a fixed $V_2$ the output switching voltage also increases thereby controlling the bias of the output.

\section{Deep Learning Simulations and Results}
\label{s5}
Simulations of the proposed architectures (DBN, SDA) were performed in MATLAB. Both generative networks with CD algorithm and behavioral model of all blocks described in section \ref{s4} were simulated. Stochastic sigmoid neuron activation and normalization circuits were simulated in Cadence Virtuoso using 90 nm CMOS design kit and Verilog-A OxRAM compact model \cite{li2015variation}.

\subsection{Stacked Denoising Autoencoder performance analysis}
\begin{figure}[t]
\centering
\includegraphics[width=\linewidth]{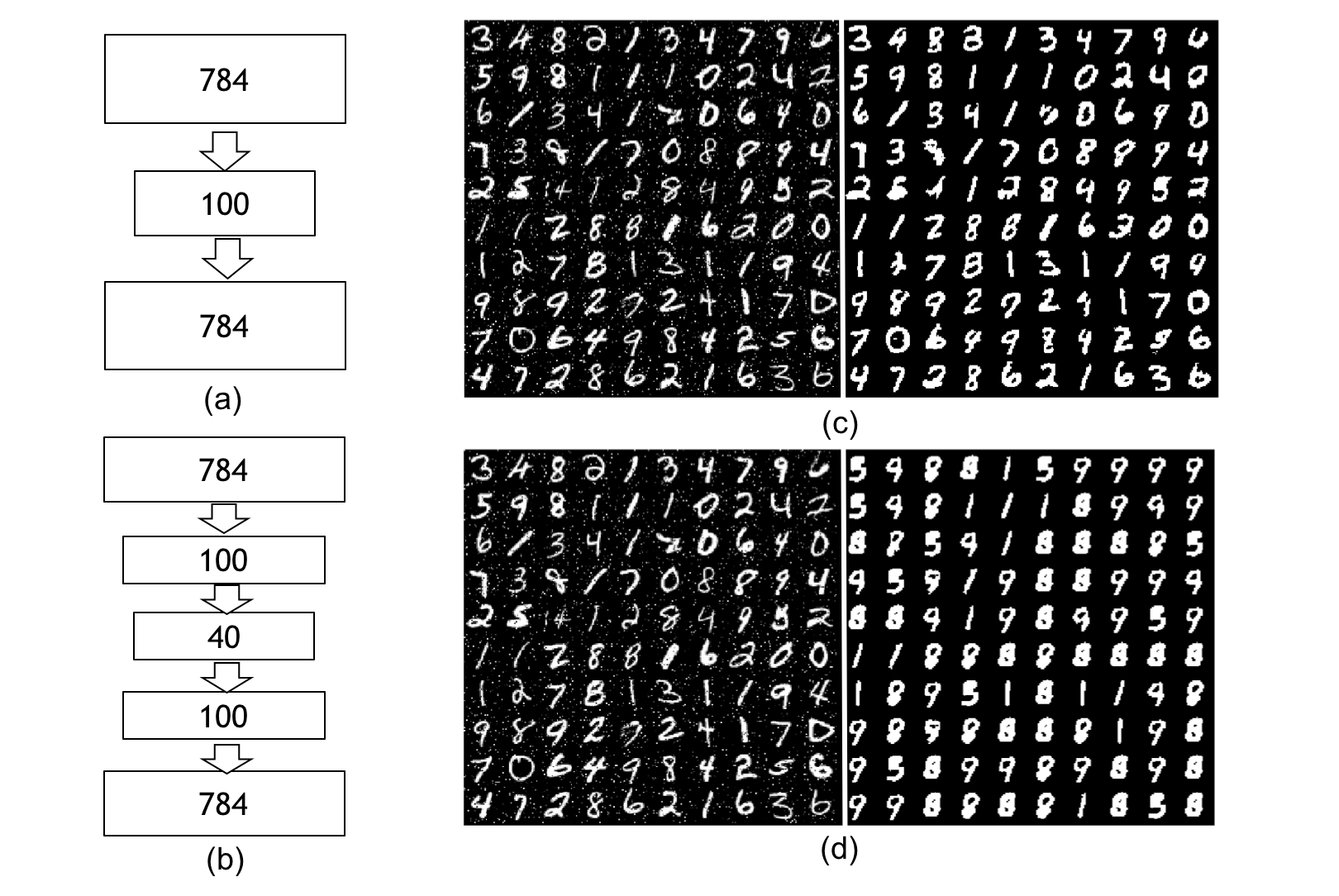}
\caption{(a) 3-layer deep SDA-1, (b) 5-layer deep SDA-2. Denoising results of 100 corrupted MNIST images for: (c) SDA1 and (d) SDA2.}
\label{fig8}
\end{figure}

We trained two autoencoder networks each having the same number of neurons in the final encoding layer, but varying levels of depth, and compared their denoising performance (see Fig. \ref{fig8}). In each network a single synaptic weight was realized using 8 OxRAM devices (8-bit resolution). All neurons have a logistic activation except for the last ten units in the classification layer, which are linear. The networks were trained on a reduced MNIST dataset of 5000 images and tested for denoising 1000 new salt-and-pepper noise corrupted images (see Fig. \ref{fig8}).

\begin{table}[!htbp]
\caption{Proposed Autoencoder performance for reduced MNIST}
\label{mnist1}
\begin{center}


    \begin{tabular}{|c|c|c|}
    \hline
    Network & Implementation & MSE \bigstrut\\
    \hline
    \multirow{2}[4]{*}{784x100x784} & Software & 0.010 \bigstrut\\
\cline{2-3}          & Hybrid OxRAM SDA & 0.003 \bigstrut\\
    \hline
    \multirow{2}[4]{*}{784x100x40x100x784} & Software & 0.049 \bigstrut\\
\cline{2-3}          & Hybrid OxRAM SDA & 1.095 \bigstrut\\
    \hline
    \end{tabular}%
\end{center}
\end{table}
Table \ref{mnist1} presents the learning performance of the proposed SDA-1 and SDA-2. Increasing depth in the network was not useful with the current learning algorithm and tuning parameters. 

\subsection{Deep Belief Network performance analysis}
\begin{figure}[t]
\centering
  \includegraphics[width=\linewidth]{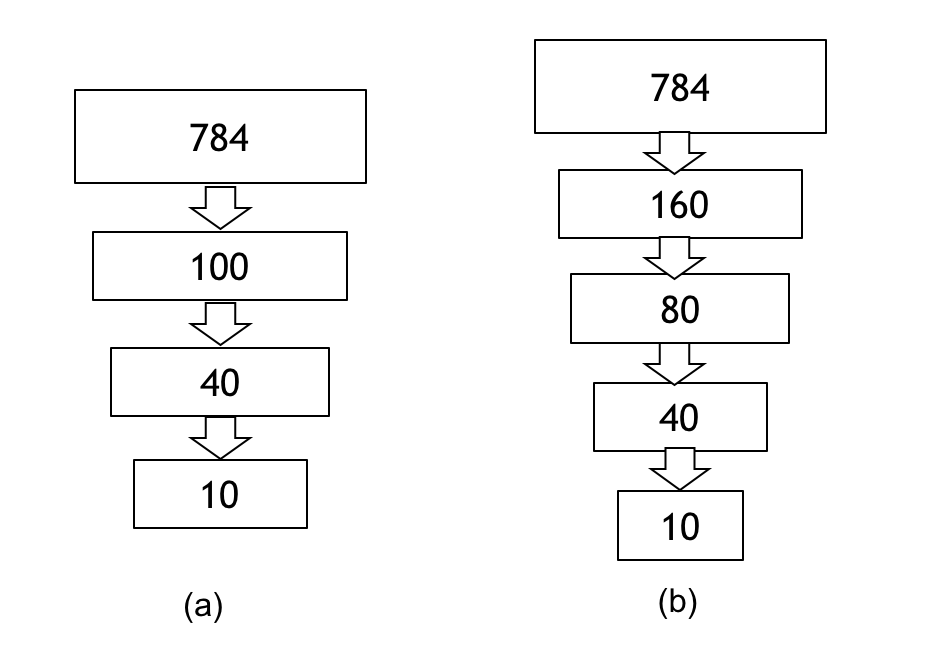}
  \caption{Simulated 4 and 5 layer DBN architecture.}
  \label{fig6b}
\end{figure}

We simulated two deep belief network architectures shown in Fig. \ref{fig6b}. (4 and 5 layer variants) Performance of the network was measured by testing on 1000 samples from the reduced MNIST dataset. The results for the same are shown in Table \ref{tab:dbn}. We measured test accuracy using 3 parameters :
\begin{enumerate}
\item Top 1 accuracy : correct class corresponds to output neuron with highest response. 
\item Top 3 accuracy : correct class corresponds to the top 3 output neurons with highest response.
\item Top 5 accuracy : correct class corresponds to the top 5 output neurons with highest response.
\end{enumerate}

From Table \ref{tab:dbn}, the performance of simulated Hybrid CMOS-OxRAM DBN matches closely with software based accuracy (2-3\% lower) for a DBN formed with 2 RBMs. There is a significant drop in test accuracy for the DBN with 3 RBMs. This is acceptable as the goal of the greedy layer-wise training is to pre-train the network to a good state before using back-propagation to allow faster convergence. Thus lower accuracy after layer-wise training for a deeper network is acceptable as the weights would be further optimized using back-propagation. 

\begin{table}[htbp]
  \centering
  \caption{Proposed DBN Performance for Reduced MNIST}
\scalebox{0.75}{
    \begin{tabular}{|c|c|c|c|c|}
    \hline
    \multirow{2}[4]{*}{Network} & \multirow{2}[4]{*}{Implementation} & \multicolumn{3}{c|}{Test accuracy} \bigstrut\\
\cline{3-5}          &       & Top-1 & Top-3 & Top-5 \bigstrut\\
    \hline
    \multirow{2}[8]{*}{784x100x40x10} & Software & 93.10\% & 98.70\% & 99.40\% \bigstrut\\
\cline{2-5}          & Hybrid OxRAM DBN & 78.70\% & 95.50\% & 98.80\% \bigstrut\\
    \hline
    \multirow{2}[8]{*}{784x160x80x40x10} & Software & 93.70\% & 98.50\% & 99.40\% \bigstrut\\
\cline{2-5}          & Hybrid OxRAM DBN & 21.30\% & 61.40\% & 79.60\% \bigstrut\\
    \hline
    \end{tabular}%
    }
  \label{tab:dbn}%
\end{table}%

\subsection{Tuning $V_{OxRAM}$ amplifying gain}
The sigmoid activation circuits in the network use a gain factor in order to balance for the low current values obtained as a result of the OxRAM device resistance values. If the amplification is low it will lead to saturation and the network will not learn a proper reconstruction of the data. This necessitates proper tuning of the amplifier gain for effective learning. In our architecture, amplifier gain for $V_{OxRAM}$ is an important hyper-parameter along with the standard ones (momentum, decay rate, learning rate, etc.) and is different for each consecutive pair of layers. A higher dimensional input to a layer will require a lower amplifying gain for $V_{OxRAM}$ and vice-versa.

\subsection{Switching activity analysis for the Proposed architecture}
\label{s6}
Resistive switching of OxRAM devices is observed in following sections of the architecture:
\begin{enumerate}
\item Synaptic matrix
\item Stochastic neuron activation
\item Internal neuron state storage. 
\end{enumerate}
RRAM devices suffer from limited cycling endurance ($\sim$ 0.1 million cycles) \cite{balatti2014pulsed}.For stochastic neuron activation, the OxRAM device is repeatedly cycled to OFF state and the voltage drop across the device is used to generate the stochastic signal fed to one of the comparator inputs. Thus the neuron activation block related switching activity depends on the number of data samples as well as number of epochs. The maximum switching per device for any layer can be estimated by using (\ref{eq1}):

\begin{equation}
\label{eq1}
N_{events} = N_{epochs} * N_{samples} * N_{batch}
\end{equation}
Another part of the architecture where the OxRAM device may observe a significant number of switching events is the synaptic matrix. Since we are interested in device endurance, we consider the worst case, i.e. the of maximum number of hits a particular OxRAM device will take during the entire weight update procedure. For worst case analysis we make the following assumptions-
\begin{itemize} 
\item While bit encoding the synaptic weight (4 or 8 or 16), there exists an OxRAM device that is switched every single time. 
\end{itemize}
Thus the maximum possible number of hits a device would take during the synaptic weight update procedure can be estimated using (\ref{eq2}): 

\begin{equation}
\label{eq2}
N_{switch events}=N_{batch}*N_{epochs}
\end{equation}
Simulated switching activity for reduced MNIST training for each neuron layer and synaptic matrix is shown in Table \ref{table:sw} and Table \ref{sw_act} corresponding to both SDA and DBN architectures respectively. Key observations can be summarized as:
\begin{itemize}
\item Increasing depth of the network increases amount of switching for hidden layers.
\item Increasing depth of the network doesn't have significant impact on the the switching events in the synaptic matrix.
\end{itemize}

\begin{table}
\centering
\caption{Maximum OxRAM switching activity for 5 layer SDA (training)} 
\begin{tabular}{|c|c|}
\hline
Device placement & Max Switching activity\\ 
\hline
L1-784 & 596\\ 
\hline
L2-100 & 3074\\ 
\hline
L3-40 & 542\\ 
\hline
W1 & 6808\\ 
\hline
W2 & 5000\\ 
\hline
\end{tabular}
\label{table:sw}
\end{table}

\begin{table}
\centering
\caption{Maximum OxRAM switching activity for 5 layer DBN (training)} 
\begin{tabular}{|c|c|}
\hline
Device placement & Max Switching activity\\ 
\hline
L1-784 & 596\\ 
\hline
L1-784 & 428 \\
\hline
L2-160 & 2069 \\
\hline
L3-80 & 3026 \\
\hline
L4-40 & 420 \\
\hline
W1 & 6798 \\
\hline
W2 & 5000 \\
\hline
W3 & 2500 \\
\hline
W4 & 2500 \\
\hline
\end{tabular}
\label{sw_act}
\end{table}

\section{Conclusion}
\label{sc}
In this paper we proposed a novel methodology to realize DGM architectures using mixed-signal type hybrid CMOS-RRAM design framework. We achieve deep generative models by proposing a strategy to stack multiple RBM blocks. 
Overall learning rule used in this study is based on greedy-layer wise learning with no back propagation which allows the network to be trained to a good pre-training stage. RRAM devices are used extensively in the proposed architecture for multiple computing and storage actions. Total RRAM requirement for the largest simulated network was 139 kB for  DBN and 169 kB for SDA.
Simulated architectures show that the performance of the proposed DGM models matches closely with software based models for 2 layers deep network. The top-3 test accuracy achieved by the DBN for reduced MNIST was $\sim$ 95.5\%. MSE of SDA network was 0.003. Endurance analysis shows resonable maximum switching activity. Future work would focus on realizing an optimal strategy to implement back-propagation with the proposed architecture to enable complete training of the DGM on the hybrid DGM architecture. 

\section*{Acknowledgement}
This research activity under the PI Prof. M. Suri is partially supported by the Department of Science \& Technology (DST), Government of India and IIT-D FIRP Grant. Authors would like to express gratitude to S. Chakraborty. The authors would like to thank F. Alibart and D. Querlioz for the HfOx device data.

\appendix
Code for all simulations discussed in the paper is available at \url{https://gitlab.com/vivekp312/oxram-sda-sim.git}. Any interested researchers can contact the authors for access to the code repository.






%
\bibliographystyle{IEEEtran}
\bibliography{mybib}

\begin{thebibliography}{10}
\providecommand{\url}[1]{#1}
\csname url@samestyle\endcsname
\providecommand{\newblock}{\relax}
\providecommand{\bibinfo}[2]{#2}
\providecommand{\BIBentrySTDinterwordspacing}{\spaceskip=0pt\relax}
\providecommand{\BIBentryALTinterwordstretchfactor}{4}
\providecommand{\BIBentryALTinterwordspacing}{\spaceskip=\fontdimen2\font plus
\BIBentryALTinterwordstretchfactor\fontdimen3\font minus
  \fontdimen4\font\relax}
\providecommand{\BIBforeignlanguage}[2]{{%
\expandafter\ifx\csname l@#1\endcsname\relax
\typeout{** WARNING: IEEEtran.bst: No hyphenation pattern has been}%
\typeout{** loaded for the language `#1'. Using the pattern for}%
\typeout{** the default language instead.}%
\else
\language=\csname l@#1\endcsname
\fi
#2}}
\providecommand{\BIBdecl}{\relax}
\BIBdecl

\bibitem{pillow2006dimensionality}
J.~W. Pillow and E.~P. Simoncelli, ``Dimensionality reduction in neural models:
  an information-theoretic generalization of spike-triggered average and
  covariance analysis,'' \emph{Journal of vision}, vol.~6, no.~4, pp. 9--9,
  2006.

\bibitem{cunningham2014dimensionality}
J.~P. Cunningham and M.~Y. Byron, ``Dimensionality reduction for large-scale
  neural recordings,'' \emph{Nature neuroscience}, vol.~17, no.~11, pp.
  1500--1509, 2014.

\bibitem{hinton1994autoencoders}
G.~E. Hinton and R.~S. Zemel, ``Autoencoders, minimum description length, and
  helmholtz free energy,'' \emph{Advances in neural information processing
  systems}, pp. 3--3, 1994.

\bibitem{bengio2007greedy}
Y.~Bengio, P.~Lamblin, D.~Popovici, and H.~Larochelle, ``Greedy layer-wise
  training of deep networks,'' in \emph{Advances in neural information
  processing systems}, 2007, pp. 153--160.

\bibitem{hot06}
G.~E. Hinton, S.~Osindero, and Y.~W. Teh, ``A fast learning algorithm for deep
  belief nets,'' \emph{Neural computation}, vol.~18, no.~7, pp. 1527--1554,
  2006.

\bibitem{raina2009large}
R.~Raina, A.~Madhavan, and A.~Y. Ng, ``Large-scale deep unsupervised learning
  using graphics processors,'' in \emph{Proceedings of the 26th annual
  international conference on machine learning}.\hskip 1em plus 0.5em minus
  0.4em\relax ACM, 2009, pp. 873--880.

\bibitem{kim2009highly}
S.~K. Kim, L.~C. McAfee, P.~L. McMahon, and K.~Olukotun, ``A highly scalable
  restricted boltzmann machine fpga implementation,'' in \emph{Field
  Programmable Logic and Applications, 2009. FPL 2009. International Conference
  on}.\hskip 1em plus 0.5em minus 0.4em\relax IEEE, 2009, pp. 367--372.

\bibitem{maaimm11}
P.~Merolla, J.~Arthur, F.~Akopyan, N.~Imam, R.~Manohar, and D.~S.~. Modha,
  ``September),'' \emph{A digital neurosynaptic core using embedded crossbar
  memory with}, vol.~45, pp. 1--4, 2011.

\bibitem{stromatias2015robustness}
E.~Stromatias, D.~Neil, M.~Pfeiffer, F.~Galluppi, S.~B. Furber, and S.-C. Liu,
  ``Robustness of spiking deep belief networks to noise and reduced bit
  precision of neuro-inspired hardware platforms,'' \emph{Frontiers in
  neuroscience}, vol.~9, 2015.

\bibitem{sqcpsvgd11}
M.~Suri, O.~Bichler, D.~Querlioz, O.~Cueto, L.~Perniola, V.~Sousa,
  D.~Vuillaume, C.~Gamrat, and B.~DeSalvo, ``Phase change memory as synapse for
  ultra-dense neuromorphic systems: Application to complex visual pattern
  extraction, electron devices meeting (iedm), 2011 ieee international, vol.,
  no,'' \emph{pp.}, vol.~4, no.~4, pp. 5--7, December 2011.

\bibitem{alibart2013pattern}
F.~Alibart, E.~Zamanidoost, and D.~B. Strukov, ``Pattern classification by
  memristive crossbar circuits using ex situ and in situ training,''
  \emph{Nature communications}, vol.~4, p. 2072, 2013.

\bibitem{yang2013memristive}
J.~J. Yang, D.~B. Strukov, and D.~R. Stewart, ``Memristive devices for
  computing,'' \emph{Nature nanotechnology}, vol.~8, no.~1, pp. 13--24, 2013.

\bibitem{de2013silicon}
B.~De~Salvo, \emph{Silicon non-volatile memories: paths of innovation}.\hskip
  1em plus 0.5em minus 0.4em\relax John Wiley \& Sons, 2013.

\bibitem{jackson2013nanoscale}
B.~L. Jackson, B.~Rajendran, G.~S. Corrado, M.~Breitwisch, G.~W. Burr,
  R.~Cheek, K.~Gopalakrishnan, S.~Raoux, C.~T. Rettner, A.~Padilla
  \emph{et~al.}, ``Nanoscale electronic synapses using phase change devices,''
  \emph{ACM Journal on Emerging Technologies in Computing Systems (JETC)},
  vol.~9, no.~2, p.~12, 2013.

\bibitem{vlzrbgkgq14}
A.~F. Vincent, J.~Larroque, W.~S. Zhao, N.~B. Romdhane, O.~Bichler, C.~Gamrat,
  J.~o.~Klein, S.~Galdin-Retailleau, and D.~Querlioz, ``Spin- transfer torque
  magnetic memory as a stochastic memristive synapse,'' \emph{In Circuits and
  Systems (ISCAS)}, vol. 2014, pp. 1074--1077, 2014.

\bibitem{wong2015memory}
H.-S.~P. Wong and S.~Salahuddin, ``Memory leads the way to better computing,''
  \emph{Nature nanotechnology}, vol.~10, no.~3, pp. 191--194, 2015.

\bibitem{burr2015experimental}
G.~W. Burr, R.~M. Shelby, S.~Sidler, C.~Di~Nolfo, J.~Jang, I.~Boybat, R.~S.
  Shenoy, P.~Narayanan, K.~Virwani, E.~U. Giacometti \emph{et~al.},
  ``Experimental demonstration and tolerancing of a large-scale neural network
  (165 000 synapses) using phase-change memory as the synaptic weight
  element,'' \emph{IEEE Transactions on Electron Devices}, vol.~62, no.~11, pp.
  3498--3507, 2015.

\bibitem{milo2016demonstration}
V.~Milo, G.~Pedretti, R.~Carboni, A.~Calderoni, N.~Ramaswamy, S.~Ambrogio, and
  D.~Ielmini, ``Demonstration of hybrid cmos/rram neural networks with spike
  time/rate-dependent plasticity,'' in \emph{Electron Devices Meeting (IEDM),
  2016 IEEE International}.\hskip 1em plus 0.5em minus 0.4em\relax IEEE, 2016,
  pp. 16--8.

\bibitem{srpj15}
A.~M. Sheri, A.~Rafique, W.~Pedrycz, and M.~Jeon, ``Contrastive divergence for
  memristor-based restricted boltzmann machine,'' \emph{Engineering
  Applications of Artificial Intelligence}, vol.~37, pp. 336--342, 2015.

\bibitem{stanford_rbm_prob}
S.~B. Eryilmaz, E.~Neftci, S.~Joshi, S.~Kim, M.~BrightSky, H.~L. Lung, C.~Lam,
  G.~Cauwenberghs, and H.~S.~P. Wong, ``Training a probabilistic graphical
  model with resistive switching electronic synapses,'' \emph{IEEE Transactions
  on Electron Devices}, vol.~63, no.~12, pp. 5004--5011, Dec 2016.

\bibitem{tnanoelm}
M.~Suri and V.~Parmar, ``Exploiting intrinsic variability of filamentary
  resistive memory for extreme learning machine architectures,'' \emph{IEEE
  Transactions on Nanotechnology}, vol.~14, no.~6, pp. 963--968, Nov 2015.

\bibitem{rramrbm}
M.~Suri, V.~Parmar, A.~Kumar, D.~Querlioz, and F.~Alibart, ``Neuromorphic
  hybrid rram-cmos rbm architecture,'' in \emph{2015 15th Non-Volatile Memory
  Technology Symposium (NVMTS)}, Oct 2015, pp. 1--6.

\bibitem{wlycwclct12}
H.-S.~P. Wong, H.-Y. Lee, S.~Yu, Y.-S. Chen, Y.~Wu, P.-S. Chen, B.~Lee, F.~T.
  Chen, and M.-J. Tsai, ``Metal--oxide rram,'' vol. 100, no.~6.\hskip 1em plus
  0.5em minus 0.4em\relax IEEE, 2012, pp. 1951--1970.

\bibitem{sqbpvvgd13}
M.~Suri, D.~Querlioz, O.~Bichler, G.~Palma, E.~Vianello, D.~Vuillaume,
  C.~Gamrat, and B.~DeSalvo, ``Bio-inspired stochastic computing using binary
  cbram synapses,'' \emph{Electron Devices, IEEE Transactions on}, vol.~60,
  no.~7, pp. 2402--2409, 2013.

\bibitem{baeumer2017subfilamentary}
C.~Baeumer, R.~Valenta, C.~Schmitz, A.~Locatelli, T.~O. Menteş, S.~P. Rogers,
  A.~Sala, N.~Raab, S.~Nemsak, M.~Shim \emph{et~al.}, ``Subfilamentary networks
  cause cycle-to-cycle variability in memristive devices,'' \emph{ACS nano},
  vol.~11, no.~7, pp. 6921--6929, 2017.

\bibitem{li2015variation}
H.~Li, Z.~Jiang, P.~Huang, Y.~Wu, H.-Y. Chen, B.~Gao, X.~Liu, J.~Kang, and
  H.-S. Wong, ``Variation-aware, reliability-emphasized design and optimization
  of rram using spice model,'' in \emph{Design, Automation \& Test in Europe
  Conference \& Exhibition (DATE), 2015}.\hskip 1em plus 0.5em minus
  0.4em\relax IEEE, 2015, pp. 1425--1430.

\bibitem{ielmini2016resistive}
D.~Ielmini, ``Resistive switching memories based on metal oxides: mechanisms,
  reliability and scaling,'' \emph{Semiconductor Science and Technology},
  vol.~31, no.~6, p. 063002, 2016.

\bibitem{bengio2009learning}
Y.~Bengio \emph{et~al.}, ``Learning deep architectures for ai,''
  \emph{Foundations and trends{\textregistered} in Machine Learning}, vol.~2,
  no.~1, pp. 1--127, 2009.

\bibitem{h10}
G.~Hinton, ``A practical guide to training restricted boltzmann machines,''
  \emph{Momentum}, vol.~9, no.~1, p. 926, 2010.

\bibitem{vincent}
H.~Vincent, P.and~Larochelle, I.~Lajoie, Y.~Bengio, and P.~Manzago, ``Stacked
  denoising autoencoders: Learning useful representations in a deep network
  with a local denoising criterion,'' \emph{Journal of Machine Learning
  Research}, no.~11, pp. 3371--3408, 2010.

\bibitem{xavier}
X.~Glorot and Y.~Bengio, ``Understanding the difficulty of training deep
  feedforward neural networks,'' \emph{In Proc. AISTATS}, vol.~9, pp. 249--256,
  2010.

\bibitem{hinton2009deep}
G.~E. Hinton, ``Deep belief networks,'' \emph{Scholarpedia}, vol.~4, no.~5, p.
  5947, 2009.

\bibitem{huang2007unsupervised}
F.~J. Huang, Y.-L. Boureau, Y.~LeCun \emph{et~al.}, ``Unsupervised learning of
  invariant feature hierarchies with applications to object recognition,'' in
  \emph{Computer Vision and Pattern Recognition, 2007. CVPR'07. IEEE Conference
  on}.\hskip 1em plus 0.5em minus 0.4em\relax IEEE, 2007, pp. 1--8.

\bibitem{sutskever2007learning}
I.~Sutskever and G.~Hinton, ``Learning multilevel distributed representations
  for high-dimensional sequences,'' in \emph{Artificial Intelligence and
  Statistics}, 2007, pp. 548--555.

\bibitem{taylor2007modeling}
G.~W. Taylor, G.~E. Hinton, and S.~T. Roweis, ``Modeling human motion using
  binary latent variables,'' in \emph{Advances in neural information processing
  systems}, 2007, pp. 1345--1352.

\bibitem{hinton2006reducing}
G.~E. Hinton and R.~R. Salakhutdinov, ``Reducing the dimensionality of data
  with neural networks,'' \emph{science}, vol. 313, no. 5786, pp. 504--507,
  2006.

\bibitem{suri5}
D.~Pan and B.~M. Wilamowski, ``A vlsi implementation of mixed-signal mode
  bipolar neuron circuitry,'' \emph{In Neural Networks, Proceedings of the
  International Joint Conference on}, vol.~2, pp. 971--976, 2003.

\bibitem{balatti2014pulsed}
S.~Balatti, S.~Ambrogio, Z.-Q. Wang, S.~Sills, A.~Calderoni, N.~Ramaswamy, and
  D.~Ielmini, ``Pulsed cycling operation and endurance failure of metal-oxide
  resistive (rram),'' in \emph{Electron Devices Meeting (IEDM), 2014 IEEE
  International}.\hskip 1em plus 0.5em minus 0.4em\relax IEEE, 2014, pp. 14--3.

\end{thebibliography}

\end{document}